# Robust pseudogap across the magnetic field driven superconductor to insulator-like transition in strongly disordered NbN films

Indranil Roy[†,1], Rini Ganguly[1], Harkirat Singh[1,2], Pratap Raychaudhuri[1]

[1]*Tata Institute of Fundamental Research, Homi Bhabha Road, Colaba, Mumbai 400005, India.*
[2] *Department of Physics, National Institute of Technology, Hazratbal, Srinagar, Jammu and Kashmir, India*

We investigate the magnetic field evolution of the superconducting state in a strongly disordered NbN thin film which exhibits a magnetic field tuned superconductor to insulator-like transition, employing low temperature scanning tunneling spectroscopy (STS). Transport measurements of the sample reveals a characteristic magnetic field, which separates the low field state where resistance decreases with decreasing temperature, i.e. $dR/dT > 0$ and a high-field state where $dR/dT < 0$. However, STS imaging of the superconducting state reveals a smooth evolution across this field and the presence of a robust pseudogap on both sides of this characteristic field. Our results suggest that the superconductor-insulator transition might be a percolative transition driven by the shrinking of superconducting fraction with magnetic field.

---

† indranil.roy@tifr.res.in

## 1. Introduction

The interplay between superconductivity and disorder is a problem of longstanding interest. Strongly disordered s-wave superconductors are recently at the focus of attention [1-5] for several noteworthy phenomena which deviate from our expectation from Bardeen-Cooper-Schrieffer (BCS) theory [6]. These include the presence of a pronounced pseudogap state above the superconducting transition temperature ($T_c$) [7, 1, 5, 8], the formation of a Cooper pair insulator [9, 10] on the insulating side of the disorder driven superconductor-insulator transition (SIT), as well as, magnetic field induced SIT [11, 12]. It has been observed in many experiments [11-13] that on application of a magnetic field, a very strongly disordered superconductor transforms into an insulator, characterized by a diverging resistance as $T \to 0$ and also shows a magnetoresistance peak, with sheet resistance as high as few Giga-Ohms in extreme cases. Even when the effect is not as dramatic the basic feature of a peak in resistance with magnetic field, is observed for a wide range of disorder close to the superconductor to insulator transition [13, 14]. The origin of this transition is still debated, but it has been suggested that the ability to tune the system continuously from one state to the other could be a manifestation of a quantum phase transition [15, 14]. Several theories such as "dirty boson" model [17-21], charge-vortex duality [22, 23], percolation based model invoking Coulomb blockade [24], have been used to explain this phenomenon.

Recently, scanning tunneling microscopy/spectroscopy (STM/S) data has provided key insights into the problem of disorder tuned SIT in absence of magnetic field [13, 25]. These studies show that in presence of strong disorder, the superconducting state becomes inhomogeneous, forming superconducting islands, of size 3-4 times the coherence length, separated by insulating patches in between [26]. The insulating regions, which increase while approaching the SIT, can be identified from the presence of a soft gap of similar magnitude to superconducting energy gap, in local density of states (LDOS) and simultaneously a suppression of coherence peaks. Therefore it is speculated that the insulating phase is also composed of Cooper pairs which lack the global phase coherence among themselves.

In this letter, we report magnetic field evolution of strongly disordered superconducting NbN thin film with $T_c \sim 1.65\ K$, using STS and transport measurements. The NbN sample, having a $k_F l \sim 1.5$ (where $k_F$ is the Fermi wave vector and $l$ is the electronic mean free path), is deep inside disordered regime [13, 27] but on the superconducting side of the SIT, such that there is already a pseudogap state present in zero magnetic field, till much above temperature than $T_c$. From the measurement of resistance ($R$) as a function of temperature ($T$) and magnetic field ($H$), we observe a magnetic field driven transition from a superconducting state, characterized by a rapid drop in $R$ at low temperatures, to an to insulator-like state where $dR/dT < 0$ down to the lowest temperature. We would like to note that despite we follow the usual nomenclature by referring to this crossing point as an SIT the high field state is not a conventional insulator with rapid divergence of resistance as observed in some systems in the very strong disorder limit. Here the high field state is closer to a bad-metal as reported earlier [28, 29]. The central result of this paper is that local density of states measured using STS smoothly evolves from the superconducting to the insulator-like state with a robust pseudogap observed on both sides of the transition.

## 2. Sample preparation and measurement

Our sample consists of an epitaxially grown NbN thin film on single-crystalline MgO (100) substrate using reactive pulsed laser deposition. A pure Nb target was ablated using a KrF excimer laser, in a $N_2$ pressure of 40 mTorr, while the substrate was kept at 750°C. The ablation was done using laser pulses of energy density of $200\ mJ/mm^2$ with a repetition rate of 4 Hz. The sample was transferred to our milli-Kelvin STM in-situ using an UHV vacuum suitcase of base pressure $10^{-10}$ Torr without exposure to air. The transport measurements were done after all STS measurements were over, using conventional four-probe technique using a current of $20\ \mu A$. The STS measurements were done employing a home-built $^3$He STM equipped with a 9 T superconducting magnet. [30] The tunneling conductance was measured by adding a $150\ \mu V$, 2 kHz ac voltage to the dc bias voltage and recording the ac response in the tunneling current using lock-in technique. Furthermore, for taking spectroscopy at a point, the tip was stabilized at that point and then the feedback loop was momentarily switched off to sweep dc bias voltage from -4 mV to +4 mV and the ac response was recorded as a function of bias voltage.

## 3. Results

We first investigate the variation of sheet resistance as a function of temperature for different magnetic fields in Fig. 1a. We observe that with increasing field the superconducting transition temperature ($T_c$) reduces and 40 kOe onwards we observe no zero-resistance state, whereas for 85 kOe we observe an insulator-like behavior, where resistance increases as temperature is reduced. Here, we define $T_c$ as the temperature where the sheet resistance falls below 0.05% of normal state resistance value. In Fig. 1b, we show the sheet resistance vs magnetic field for different temperatures. As we increase temperature, as expected, the magnetoresistance curves cross each other within the magnetic field range of 74 kOe to 84 kOe. However the crossing field in the temperature range of our measurement is not universal; it gradually decreases with increase in temperature, similar to that observed in moderately disordered $InO_x$ films [14]. This non-universal crossing precludes the possibility of scaling of the data around a single crossover field [31-33]. We also observe here for temperatures below 1.65 K, the R-H curves have reached a maxima at around 100 kOe, denoted as $H_{max}$, above which the curves start to decrease slightly. We call this maxima the 'magnetoresistance peak'.

To inspect the inhomogeneity in superconducting order parameter we acquire area spectroscopy of 200 nm x 200 nm area on a 32x32 pixel grid for zero field at 350 mK. The sample in consideration, being in a highly disordered regime, has a temperature independent V-shaped background in the spectra extending up to high bias, arising from Altshuler-Aronov type electron-electron interactions. [25] Unlike clean NbN case where the spectrum is fully gapped, the spectra in strong disordered samples such as this, show a substantial Zero Bias Conductance (ZBC) value. Furthermore, the height of coherence peaks differ significantly over the space. For example, Fig. 2a shows a spectrum (Spectrum-I) with significant coherence peak, a spectrum (Spectrum-II) without that obtained at a different point and a V-shaped spectrum (Spectra-III) taken at 10 K for zero field which is similar for all points and will be used to normalize all spectra to remove the V-shaped background. Fig. 2b shows the normalized spectra obtained by dividing all spectra by spectrum-III of Fig. 2a. From this point all spectra or conductance maps shown will be normalized in this fashion. [25] The normalized ZBC ($G_N(0)$) map shown in Fig. 2c shows fragments and islands of higher and lower values of $G_N(0)$, signifying the inhomogeneity of order parameter over space. [26]

To understand the role of variation of order parameter in SIT, we investigate the area spectroscopy over the same area for different applied magnetic fields. Each of the normalized ZBC maps obtained for different fields, are shown in Fig. 3a. As shown in Fig. 3b, the histogram obtained from these ZBC maps also remain almost similar in width till 60 kOe field, and starts to widen from 75 kOe and in 85 kOe the histogram widens significantly. Standard deviations of the histograms of the in-field ZBC maps increase slowly with increasing magnetic field, shown in Fig. 4. We can understand this in terms of fragmentation of islands further into smaller islands of higher and lower values of $G_N(0)$.

We plot in Fig. 5a the spectra for different fields obtained through averaging spectra for each fields over the 200 nm x 200 nm area at 350 mK. Though we see small change in zero bias conductance, the basic feature in the average spectra remains unchanged for different fields up to 85 kOe. In Fig. 5b, we explore temperature dependence of averaged superconducting gap for different fields with the R-T data superposed on it. This average is done on spectra acquired on 128 points along lines passing through a vortex patch, as well as, a superconducting island [34]. We mark two points on such maps on the temperature axis, one where the sheet resistance falls off to 0.05% of its normal state value, as $T_c$ and the other point where $G_N(0)$ falls off to 95% of its normal state value, as $T^*$. We plot these two points for different fields on H-T parameter space in Fig. 5c and observe that $T^*$ remains almost constant at about 7-8 K for different fields, whereas, $T_c$ decreases rapidly with increasing magnetic field. The parameter space between the $T^*$ line and the $T_c$ line is called the pseudogap region.

## 4. Discussion and conclusion

The robustness of this pseudogap in all magnetic fields gives us one key piece to the puzzle of SIT. The electronic information obtained from spectra of superconducting state, as well as, the insulator-like state proves that the two states are fundamentally same, characterized only by different volume fraction of superconducting regions and average size of superconducting puddles. Based on these observations we can give a physical picture of the SIT. The superconducting state in presence of strong disorder fragments into superconducting islands which have different phases of order parameter; among these islands which are Josephson coupled, go into zero-resistance state via Andreev tunneling of Cooper pairs. [24, 26] When an external magnetic field is applied, the superconducting islands further fragment into smaller islets separated by patches of vortices. When magnetic field is increased the sizes of these islands decrease due to increased number of lines of vortices. The Josephson coupling between such islands is weakened, and at a characteristic field the global phase coherence between the islands is lost. This corresponds to the line $T_c(H)$ where the zero resistance state is destroyed. Above this field the superconductor is simply a disordered network of superconducting puddles, phase incoherent with each other, embedded in an insulating matrix.

To understand the *R-H* behavior above the destruction of the zero resistance state we can use a simple resistance network model. Here we simulate effective resistance from the $G_N(0)$ maps of Fig. 3a, exactly following the prescription of Ref. [24]. We first convert our $G_N(0)$ maps into binary maps (Fig. 6a-b) using a thumb-rule: points with $G_N(0)$ values lower than a particular value (which we adjust at $G_N(0)$=0.52 to obtain the best fit with the data) are considered to be superconducting points, whereas, points with $G_N(0)$ values higher than that, are considered to be insulating. Consequently we assign three types of resistance values for three types of junctions: Superconductor-Superconductor junctions ($R_1$), Superconductor-Normal junctions ($R_2$) and Normal-Normal junctions ($R_3$). We have taken $R_1 \ll (R_2, R_3)$, but nonzero, to avoid numerical runaways. $R_2$ on the other hand should depend on the charging energies of the superconducting islands. Following ref. 24, we have assumed charging energy, consequently $R_2$ to be equal for all Superconductor-Normal junctions. By using this prescription, we obtain a resistor network as is shown in Fig. 6c. The equivalent resistance of this resistor network is calculated from two nodes lying on the opposite edges [35]. All possible combinations are averaged over to get the final value of the resistance in arbitrary unit of $R_0$, obtained at 350 mK for different magnetic fields, which is plotted in Fig. 6d and is compared with 350 mK *R-H* data. Though the model is oversimplified, it still captures very well the key feature of the data, which is the increase in the resistance of the system as the superconductor fragments into smaller islands.

Now, one can extend the same model to understand what happens near the magnetoresistance peak and beyond. Though we have taken charging energy of the superconducting islands to be independent of their size in our simulation, actually charging energy is inversely proportional to the size of the islands, and hence the Cooper pairs tunneling via Andreev tunneling have to overcome this charging energy [24]. Due to the presence of the charging energy, there are two channels for transport, one via Andreev tunneling through the superconducting islands, giving a resistance, $R_S$ due to the charging energy, and transport via normal patches, giving a resistance, $R_I$. Below the magnetoresistance peak, $R_S < R_I$ and the tunneling path through the superconducting islands is preferred. As magnetic field is increased the superconducting islands become smaller and as a result their charging energy, and in turn the resistance, $R_S$ increases. At a higher magnetic field, $H = H_{max}$, the two channels are equivalently resistive, i.e., $R_S = R_I$, which gives rise to the magnetoresistance peak. At an even higher magnetic field, i.e., $H > H_{max}$, the superconducting islands acquire a size where transport through the superconducting islands is energetically unfavorable due to Coulomb blockade and transport occurs through the normal channels. Therefore as the magnetic field is increased the decrease in the size of superconducting islands results increase in the number of normal channels thereby decreasing the resistance. We observe a signature of this decrease close to 110 kOe.

There are reports of a two-step magnetic-field induced SIT [16], which observe a second SIT at a lower magnetic field. Ref. [16] invokes evolution of vortex state to explain this second SIT. We want to make a remark about our system here, which is so disordered that we do not observe a vortex lattice [26], hence we can safely conclude that even if such a second SIT at a lower magnetic field exists, we have not observed it owing to the much disordered vortex state already present in our system. In which case, we are dealing with the SIT present at a higher magnetic field as per Ref. [16].

In summary, our results trigger the idea that in the SIT, both superconducting and insulating states are intrinsically similar. The emerging picture from our measurement is that the shrinking of superconducting fraction with emergent granularity because of disorder, as well as, magnetic field, gives rise to a percolative transition from a state with dR/dT > 0 to a state where dR/dT <0. These results seriously question the description of the magnetic field driven SIT as a manifestation of an underlying quantum phase transition. Whether this picture would fundamentally change for samples with $k_Fl <<1$ is at the moment an open question. We believe that this issue should be further investigated on more strongly disordered samples, at lower temperatures and higher magnetic fields, which exhibit a large magnetoresistance peak.

*Acknowledgement*: The authors would like to thank John Jesudasan and Vivas Bagwe for help with experiments. This work was funded by the Department of Atomic Energy, Govt. of India and the Department of Science and Technology, Govt. of India (Grant No. EMR/2015/000083).

*Author contributions*: IR and RG performed the STS measurements. IR analyzed the data. HS prepared the sample and performed the transport measurements. PR conceived the problem and supervised the project. All authors read the manuscript and commented on the paper.

---

**Figure Captions**

**Figure 1**

(a) Sheet resistance vs temperature for different magnetic fields. Thinner to thicker lines denoting increasing magnetic fields, starting with the dashed line for 0 kOe. Zero-resistance state in lowest temperature is obtained up to 30 kOe. For 85 kOe the curve shows insulator-like $dR/dT < 0$ down to lowest temperature. The normal state of the sample has a negative slope of resistance, characteristic feature of a strongly disordered scenario. (b) Sheet resistance vs. magnetic field for different temperatures. Dashed line for 350 mK and after that thinner to thicker lines denoting increasing temperatures. Peak of the magnetoresistance is observed at $H_{max} \sim 100\ kOe$ for temperatures below 1.65 K.

**Figure 2**

(a) Three spectra are shown here: (I) spectrum with significant coherence peak at 350 mK; (II) spectrum without coherence peak at 350 mK; (III) spectrum taken at 10 K at zero field for normalization. Spectra I and II are taken at two different points on the same area of the sample. Whereas Spectrum III is similar for all points at 10 K. (b) Normalized spectrum (I) with coherence peak; (II) without coherence peak and (III) flat spectrum with normalized conductance value 1 for all bias. (c) 200 nm x 200 nm map taken as 32 pixels x 32 pixels of normalized conductance values taken at zero bias at 350 mK. The map shows islands of higher and lower values of $G_N(0)$, showing variation in superconducting order parameter.

**Figure 3**

(a) $G_N(0)$ maps for magnetic fields 0 kOe, 20 kOe, 40 kOe, 60 kOe, 75 kOe and 85 kOe respectively at 350 mK. (b) Histograms of $G_N(0)$ for magnetic fields 0 kOe, 20 kOe, 40 kOe, 60 kOe, 75 kOe and 85 kOe respectively at 350 mK. The histograms for 0 kOe to 40 kOe are almost similar, whereas for much higher fields, 60 kOe, 75 kOe and 85 kOe the histograms have a progressively shorter peak, as well as larger width, signifying larger number of fragmentations for higher fields.

**Figure 4**

The connected dots correspond to standard deviation of zero bias conductance histograms as a function of magnetic field, obtained from Fig. 3b. The connecting line is a guide to the eye.

**Figure 5**

(a) Normalized average spectra for different magnetic fields show continuous change of $G_N(0)$ values with increasing magnetic field. Thinner to thicker lines correspond to lower to higher magnetic fields. (b) Temperature variation of gap for magnetic fields 0 kOe, 20 kOe, 40 kOe, 60 kOe, 75 kOe and 85 kOe respectively, along with in-field sheet resistance as a function of temperature superimposed onto it. The temperature where sheet resistance drops to 0.05% of its normal state value is marked as $T_c$, which is absent for and above 40 kOe. The temperature where $G_N(0)$ drops to 95% of its normal state value is marked as $T^*$. (c) $T_c$ and $T^*$ are plotted on H-T parameter space. We observe that, while $T_c$ decreases rapidly with increasing magnetic field, $T^*$ remains almost constant. Comparing gap variation with temperature for different magnetic fields we observe that the pseudogap feature is omnipresent in all magnetic fields.

**Figure 6**

(a)-(b) Binary version of G_N (0) maps at 350 mK for 40 kOe and 85 kOe respectively. (c) A schematic of the resistor network obtained from binary maps. The grey squares represent superconducting points, while the white ones are insulating points. Each pixel in the binary map in (a)-(b) correspond to a node in a resistor network as in (c). The resistance values in the network are denoted as Superconductor-Superconductor junction: $R_1$, Superconductor-Normal junction: $R_2$ and Normal-Normal junction: $R_3$. (d) The percolative simulation based on $G_N(0)$ maps from Fig. 3b gives Resistance in units of arbitrary $R_0$ as a function of magnetic field (red squares), which is compared with 350 mK *R-H* data (black line).

**Figure 1**

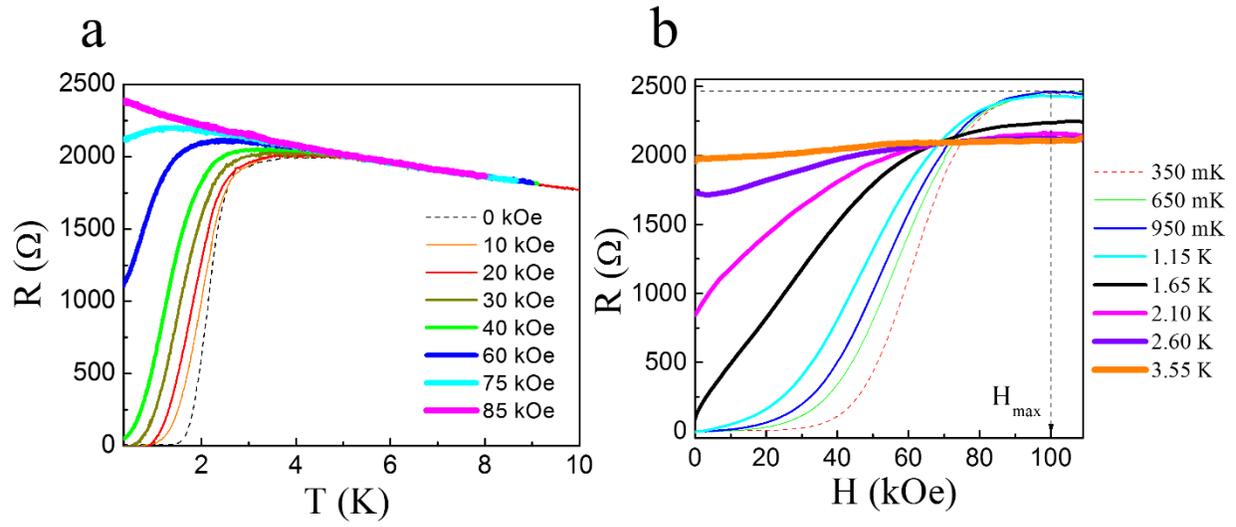

**Figure 2**

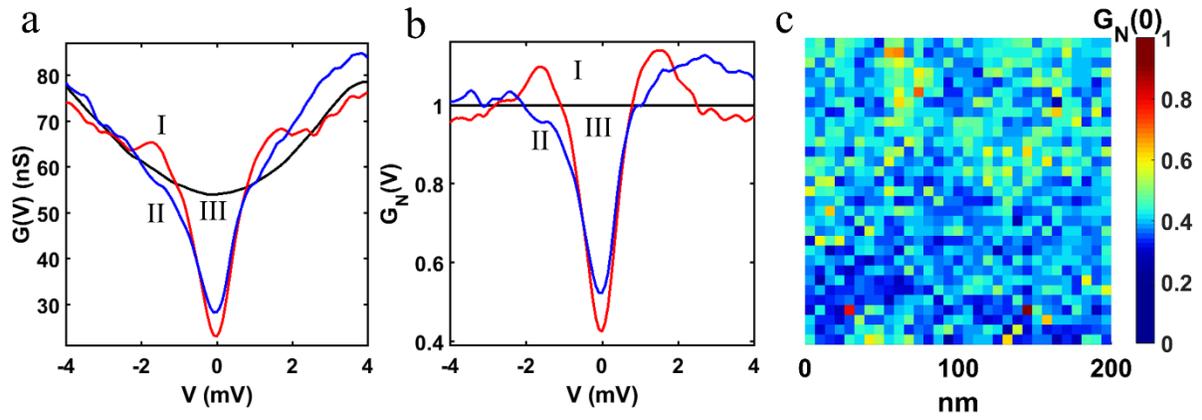

**Figure 3**

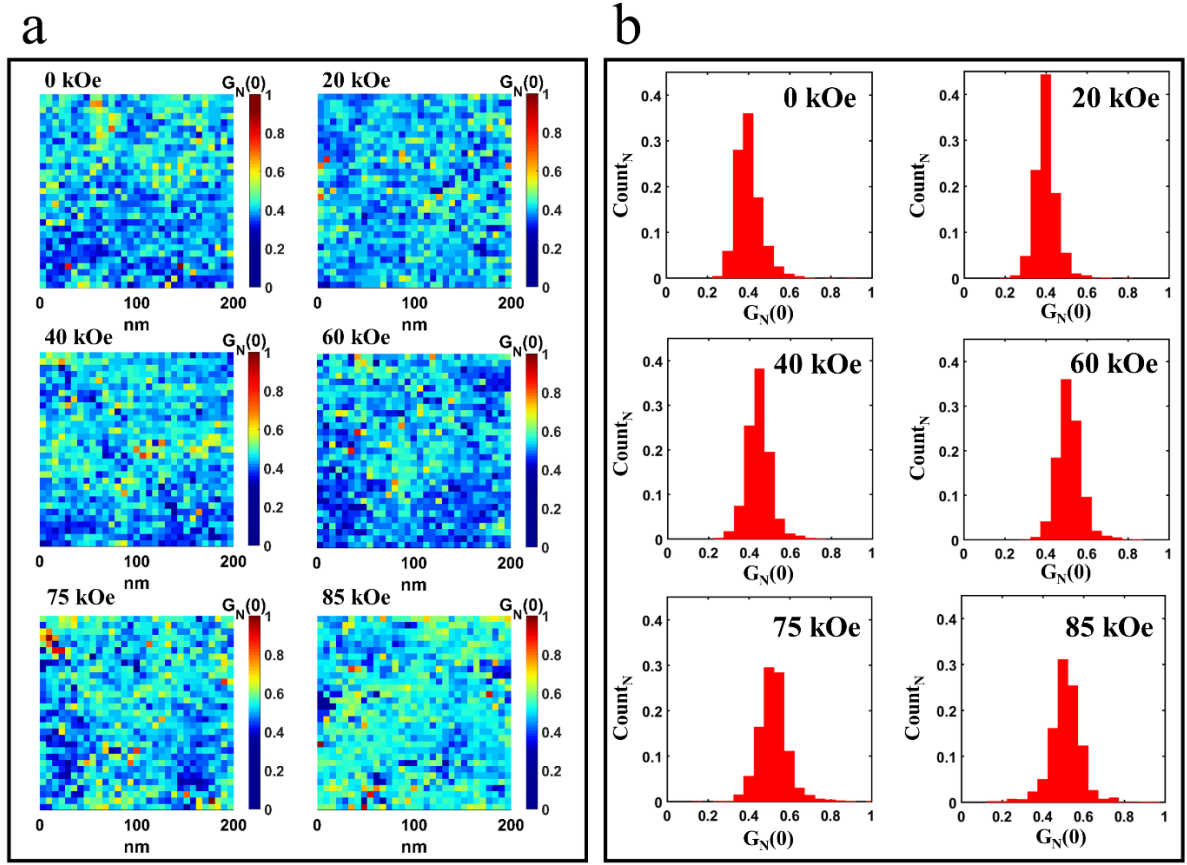

**Figure 4**

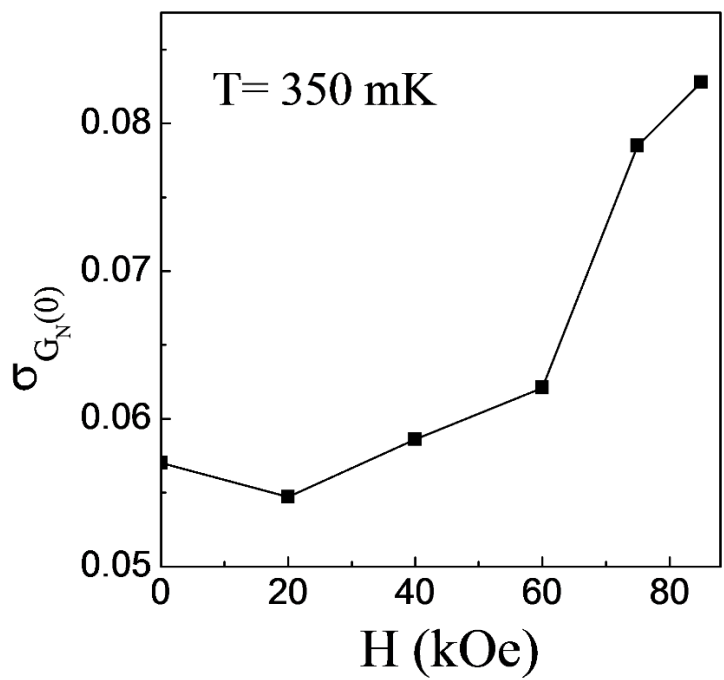

**Figure 5**

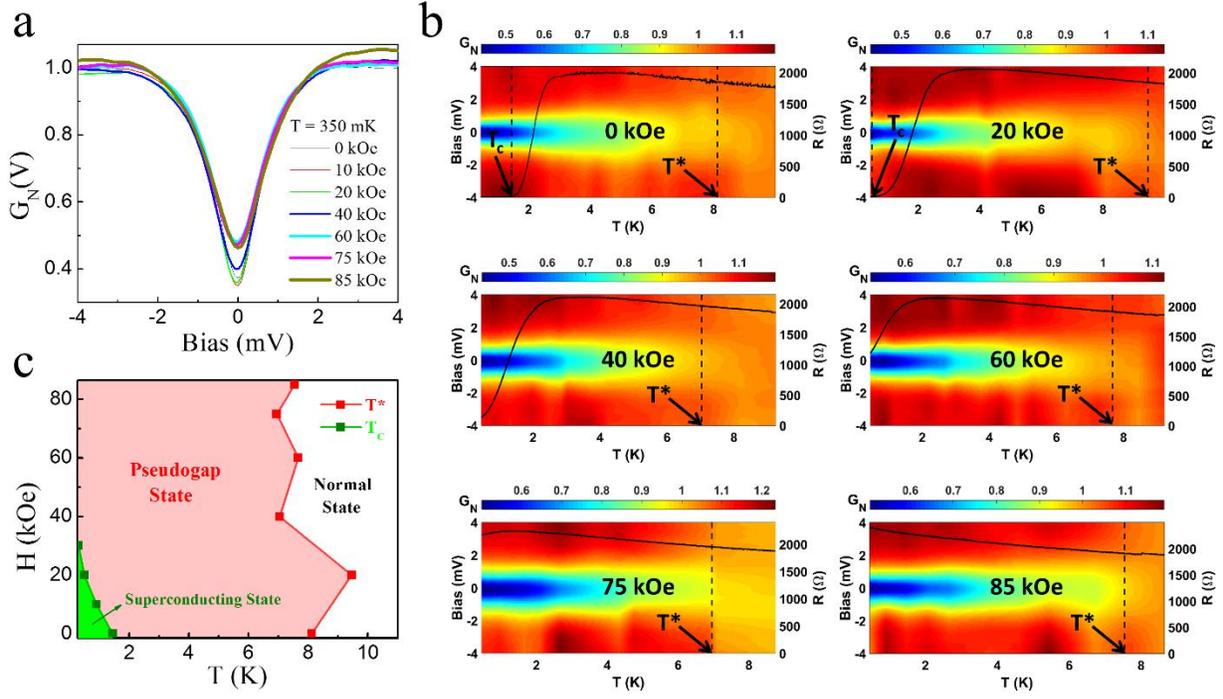

**Figure 6**

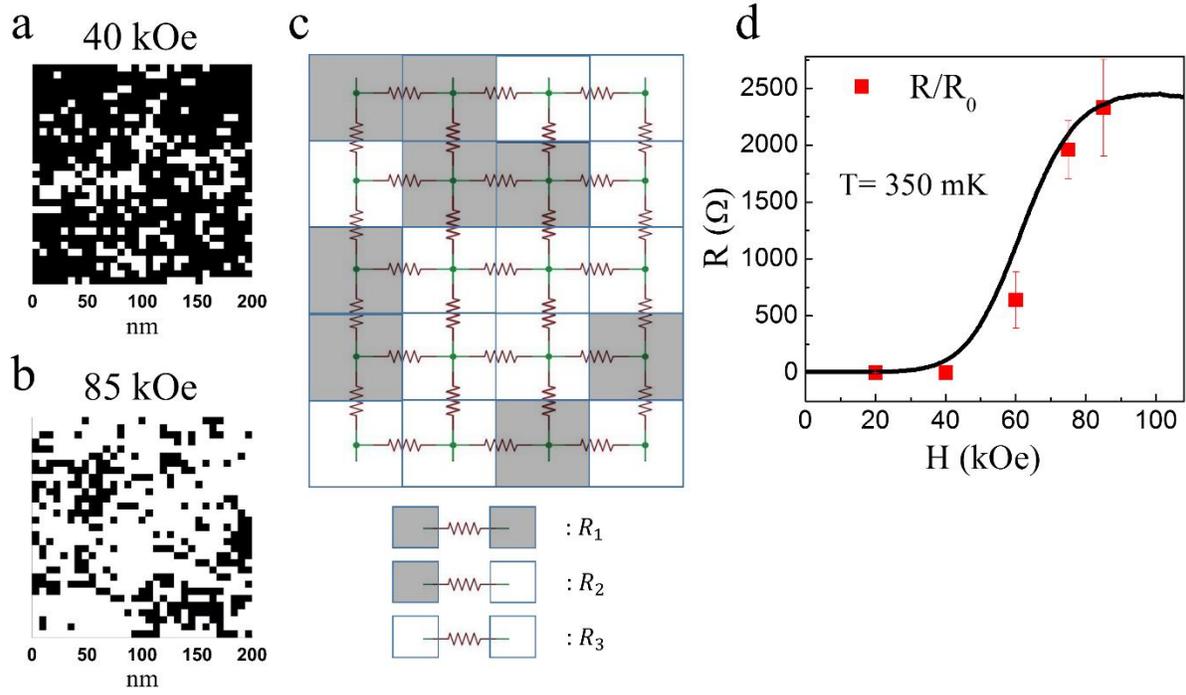

# Supplementary material

# Robust pseudogap across the magnetic field driven superconductor to insulator like transition in strongly disordered NbN films

Indranil Roy[1], Rini Ganguly[1], Harkirat Singh[1], Pratap Raychaudhuri[1]

[1]*Tata Institute of Fundamental Research, Homi Bhabha Road, Colaba, Mumbai 400005, India.*

**Section 1.** We present here comparison between $G_N(0)$ maps and conductance maps taken at bias= 0.8 mV. We observe a slightly negative cross-correlation between these two maps, which weakens as the magnetic field is increased. In Fig. s1, we present comparison between these two maps for 20 kOe, 40 kOe and 60 kOe at 350 mK. For these three fields the cross-correlations are -0.1017, -0.0707 and -0.0096 respectively. The cross-correlation histograms too show a slightly negative slope with a large scatter. The cross-correlations are weaker here because of larger disorder in the system. Similar calculations for weak disorder have been reported previously showing similar cross-correlations. [1]

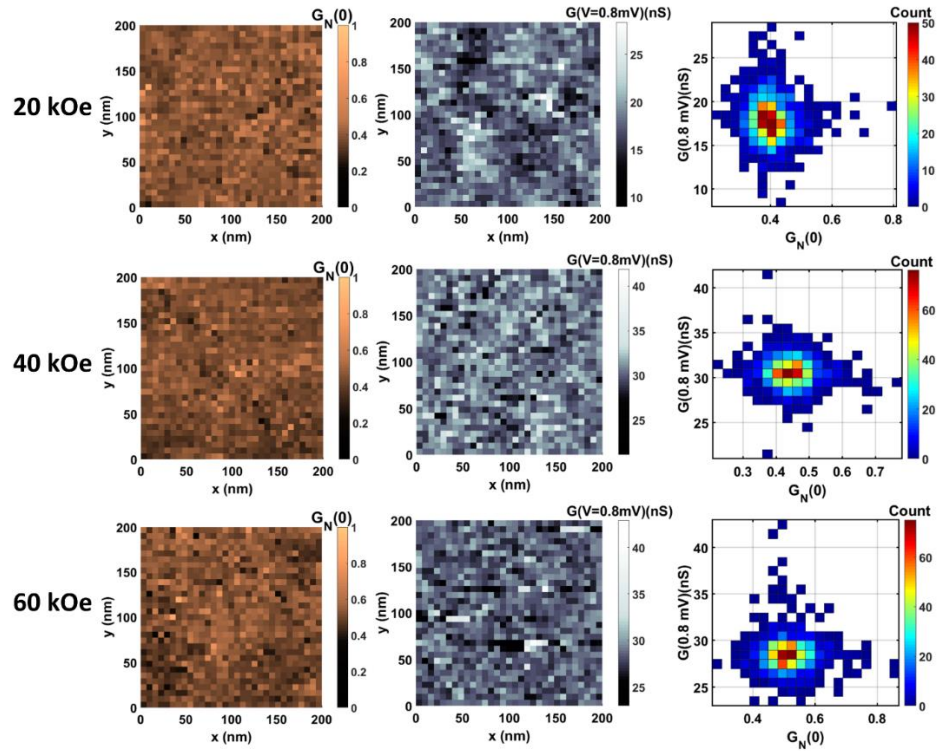

*Figure s1*: The leftmost column shows $G_N(0)$ maps for 20 kOe, 40 kOe and 60 kOe respectively at 350 mK. The middle column shows conductance maps (in nS) at bias (V) = 0.8 mV for 20 kOe, 40 kOe and 60 kOe respectively. The cross-correlations are -0.1017, -0.0707 and -0.0096 respectively for these three fields. The rightmost column shows the cross-correlation histograms for these three fields. These histograms have a slightly negative slope with a large scatter.

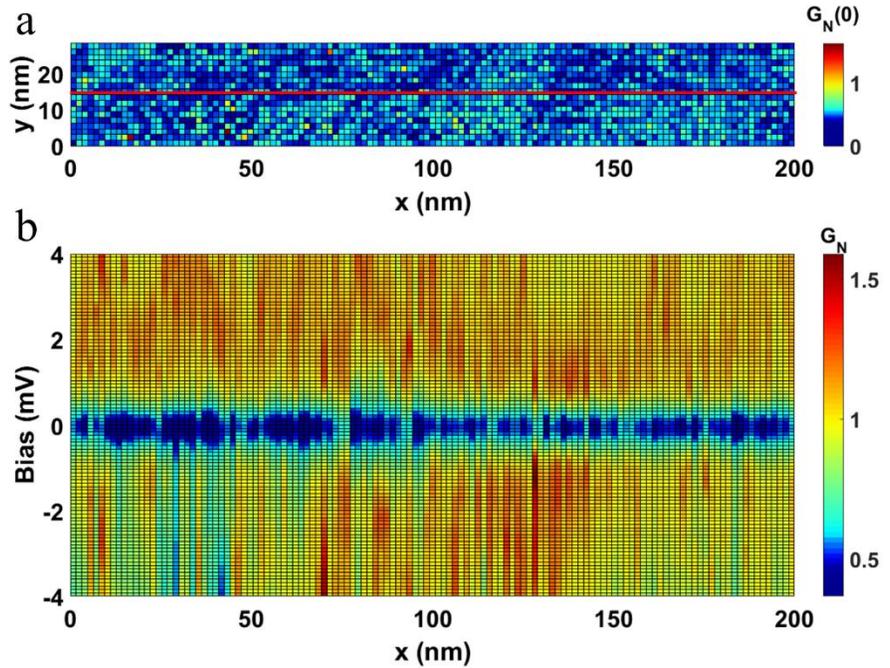

**Section 2.** We present in Fig. s2a $G_N(0)$ map obtained from area spectroscopy for 20 kOe field taken on a 128 pixels grid on 200 nm x 28 nm area. On a $G_N(0)$ map larger values of $G_N(0)$ represent cores of vortices and smaller values represent superconducting regions. We choose a line passing through lines of vortices and check individual spectra of each points along this line. Such a line is shown in Fig. s2b where all $G_N(0)$ values are much less than unity, signifying gapped states in all points, be it inside or outside vortices.

*Figure s2*: (a) $G_N(0)$ map obtained from area spectroscopy for 20 kOe field taken on a 128 pixels grid on 200 nm x 28 nm area at 350 mK. The red line is a representative line passing through vortices. (b) Spectra along the red line of Fig. s2a. All spectra have $G_N(0)$ values much less than unity.

---

1. Rini Ganguly and Indranil Roy et al, Phys. Rev. B 96, 054509 (2017).